\documentclass[prd,showpacs,twocolumn,showkeys,preprintnumbers,%
               nofootinbib,letterpaper,floatfix]{revtex4}
\usepackage{graphicx}                                 
\usepackage{amsmath,amsfonts,amssymb}
\graphicspath{{./Figures/}}                           
\usepackage{graphicx}
\newcommand{\coul}{\mathrm{Coul}}
\newcommand{\wilson}{\mathrm{Wilson}}
\newcommand{\mxx}{\mathrm{max}}
\newcommand{\Fig}[1]{Fig.~\ref{#1}}
\newcommand{\Tab}[1]{Table~\ref{#1}}

\newcommand{\Eq}[1]{Eq.~(\ref{#1})}
\renewcommand{\Re}{\operatorname{\mathfrak{Re}}}
\newcommand{\Tr}{\operatorname{Tr}}
\newcommand{\Ncp}{N_{\mathrm{cp}}}

\begin{document}
\preprint{HU-EP-08/04, ADP-08-01/T661}

\title{The effective Coulomb potential in SU(3) lattice Yang-Mills theory}
\author{A.~Voigt\footnote{present address: Max-Planck-Institut f\"ur Meteorologie, 
D-20146 Hamburg, Germany}, E.-M.~Ilgenfritz and M.~M\"uller-Preussker }
\affiliation{Humboldt-Universit\"at zu Berlin, Institut f\"ur Physik,
D-12489 Berlin, Germany}
\author{A.~Sternbeck}
\affiliation{CSSM, School of Chemistry \& Physics, The University of
  Adelaide, SA 5005, Australia} 

\date{June 2, 2008}

\begin{abstract}
  We study the infrared behavior of the effective Coulomb potential in
  lattice $SU(3)$ Yang-Mills theory in the Coulomb gauge. We use 
  lattices up to a size of $48^4$ and three values of the inverse coupling,
  $\beta=5.8, 6.0$ and $6.2$. While finite-volume effects are hardly
  visible in the effective Coulomb potential, scaling violations and a
  strong dependence on the choice of Gribov copy are observed.
  We obtain bounds for the Coulomb string tension that are in agreement with 
  Zwanziger's inequality relating the Coulomb string tension to the Wilson
  string tension.
\end{abstract}

\keywords{Coulomb gauge, Coulomb potential, Gribov problem, simulated
  annealing, running coupling, ghost propagator}
\pacs{11.15.Ha, 12.38.Gc, 12.38.Aw}
\maketitle

\section{Introduction}
\label{sec:introduction}

Confinement is one of the peculiar features of Quantum Chromodynamics,
the theory of strong interactions. Thanks to 25 years of intensive
research in the field of lattice gauge theory, a few mechanisms for 
confinement have been identified. These mechanisms are associated 
either with monopoles or vortices and seem to be closely related to 
each other~\cite{Greensite:2003bk}. In this context, confinement was
and is stated mostly by a non-vanishing string tension
$\sigma_{\wilson}$ which expresses the minimal energy of the gluon field
between a pair of static quarks. The string tension is defined by
Wilson loops and can be extracted in the limit of large Euclidean time
from the Wilson loop's area-law decay. This definition, however, is
not completely satisfying, because not only quarks but also gluons are
confined, and there is no area law in the more realistic case of light
dynamical quarks present in the vacuum.

There are two other, though less popular, approaches that might help
to shed additional light on the phenomenon of confinement. One is given
by the Hamiltonian approach which promises to present an understanding
not only of bound states but also of the vacuum structure in terms of
wave functionals. The other is a more field-theoretically inspired
approach that focuses on the QCD Green's functions and their infrared
behavior. The QCD Green's functions may serve as input to a hadron
phenomenology based on the Bethe-Salpeter and Faddeev equations. 
There, the ultimate goal is to attain a coherent description
of hadronic states and processes based on the dynamics of confined
gluons, ghosts and quarks (see, e.g., Ref.~\cite{Alkofer:2000wg}).

Both the Hamiltonian approach and investigations of QCD Green's functions 
require to fix a gauge. This introduces the well-known Gribov
ambiguity present in the Coulomb as well as in covariant gauges. One
should keep in mind that the confinement mechanisms associated with
monopoles and vortices, that received credit by reproducing the
Wilson string tension, also mostly require a gauge condition.
In the Coulomb gauge, the Gribov ambiguity represents a severe 
source of uncertainty and its effect on the results must be faithfully
checked. On the other hand, the Coulomb gauge yields a particularly
interesting confinement picture called the Gribov-Zwanziger
scenario~\cite{Gribov:1977wm,Zwanziger:1998ez}. 
This scenario might provide an understanding of confinement 
even in the presence of dynamical quarks when the Wilson-loop 
criterion fails.

A central element of the Gribov-Zwanziger confinement scenario in 
Coulomb gauge is the instantaneous color-Coulomb potential involving 
the Faddeev-Popov operator $M$ (in Coulomb gauge) and the infrared
spectral properties of the latter~\cite{Greensite:2004ur,Nakagawa:2007fa}.
The expression
\begin{equation}
  V_{\coul} (x-y)\delta^{ab} = \left\langle
    g^2\left[M^{-1}(-\triangle)M^{-1} \right]^{ab}(x,y) \right\rangle \; 
\label{eq:coulomb_potential}
\end{equation}
is defined through the vacuum expectation value of the potential part of the
Hamilton operator
\begin{equation}
  H  =  \frac{1}{2} \int d^{3}x \, \left( {\vec \Pi}_{tr}^{2}({\vec x})
    + {\vec B}^{2}({\vec x}) \right) + H_{\coul}
  \label{eq:Hamiltonian}
\end{equation}
resulting from the elimination of longitudinal degrees of freedom.
Here the potential term $H_{\coul}$ is expressed in terms of the color charge
density (including external sources and the charge density of the gluon field
itself) by means of the color-Coulomb potential, 
\begin{equation}
  H_{\coul}  =  \frac{1}{2} \int d^{3}x d^{3}y \, \rho^{a}({\vec x})
  V_{\coul} (x-y)\delta^{ab}
  \rho^{b}({\vec y}) \; .
  \label{eq:Coulomb_energy}
\end{equation}
As Zwanziger has shown~\cite{Zwanziger:1998ez}, the Coulomb potential does
not equal the Wilson potential $V_{\wilson}$ used to extract the string tension
$\sigma_{\wilson}$ as an order parameter for confinement. Instead, for large
spatial distances $r$ the Coulomb potential represents an upper bound for the
rise of the Wilson potential, 
\begin{equation}
  V_{\wilson}(r) \leq - \frac{4}{3} V_{\coul}(r) \; .
  \label{eq:Zwanziger_inequality}
\end{equation}
In other words, there is no confinement without Coulomb confinement
since the Coulomb string tension is an upper bound for the Wilson string 
tension~\cite{Zwanziger:2002sh}, 
\begin{equation}
  \sigma_{\wilson} \leq \frac{4}{3} \sigma_{\coul} \; .
  \label{eq:Zwanziger_inequality_2}
\end{equation}
Zwanziger has continuously developed the confinement scenario originally 
proposed by Gribov~\cite{Gribov:1977wm}. He has put forward the Coulomb 
potential as a new order parameter for 
confinement~\cite{Zwanziger:1998ez,Zwanziger:2002sh,Zwanziger:2003de}.
In fact, the Coulomb potential is expected to linearly rise at large~$r$ even in
the presence of dynamical quarks when the Wilson-loop criterion fails. Recent
lattice studies have shown, however, that the relation
(\ref{eq:Zwanziger_inequality}) is only a {\it necessary}~\cite{Greensite:2003nf}
condition for confinement, and that the Coulomb potential 
can be linearly rising with spatial distance even in the deconfinement
phase~\cite{Nakagawa:2007fa,Nakagawa:2006fk}.

Using lattice techniques, a linearly rising Coulomb
potential~\cite{Greensite:2003nf,Nakamura:2005ux,Nakagawa:2006fk}
and a connection between the center-vortex mechanism and the Gribov-Zwanziger
scenario~\cite{Greensite:2004cz,Greensite:2004ke,Greensite:2004ur}  
have been observed. Furthermore, Greensite et al.\ proposed~\cite{Greensite:2003xf} 
to use correlators of partial Polyakov loops to measure the Coulomb potential. 
Corresponding $SU(2)$ as well as $SU(3)$ studies revealed that the Coulomb
string tension $\sigma_{\coul}$ could well be 2-3 times larger than
the Wilson string tension
$\sigma_{\wilson}$~\cite{Greensite:2003xf,Greensite:2004cz,Nakamura:2005ux,Nakagawa:2006fk}. 
This is in contrast to results of $SU(2)$ studies where the Coulomb potential 
was measured by means of its very definition via \Eq{eq:coulomb_potential} suggesting
$\sigma_{\coul} = \sigma_{\wilson}$ \cite{Cucchieri:2002su,Langfeld:2004qs}.  
In the present study we provide a (yet missing) thorough measurement
of the Coulomb potential in $SU(3)$ gauge theory based on its very
definition in \Eq{eq:coulomb_potential}. We investigate the relation
between $\sigma_{\wilson}$ and $\sigma_{\coul}$ and find, though
hedged with large uncertainty, $\sigma_{\coul}$ to be 1.6 times larger than
$\sigma_{\wilson}$. The origin of the systematic uncertainty will be discussed. 

The paper is organized as follows. In Sect.~\ref{sec:detailsofsimulation} 
we describe the details of our numerical simulation and define the 
lattice observables measured. We investigate finite-volume effects, 
lattice-spacing effects and the effects due to the Gribov ambiguity 
in Sect.~\ref{sec:systematics}. In Sect.~\ref{sec:infrared}  
we analyze the infrared behavior of the effective Coulomb potential. 
A summary concludes this paper.

\section{Details of the numerical simulation}
\label{sec:detailsofsimulation}

\subsection{Lattice samples and gauge-fixing algorithms}
\label{sec:gaugefixing}

For our study we use the standard lattice formulation of $SU(3)$
Yang-Mills theory in Coulomb gauge where we always start from
non-gauge-fixed $SU(3)$ gauge configurations and apply the Coulomb gauge
condition subsequently. Our sets of gauge configurations were generated
with Wilson's one-plaquette action at three values of the inverse
coupling, $\beta = 5.8$, $6.0$ and $6.2$, for a couple of lattice sizes
$L_{s}^{3} \times L_{t}$ where $L_{t}$ and $L_{s}$ denote the
spatial and temporal lattice extension, respectively. We have only
considered hyper-cubic lattices with $L_{s}=L_{t}=L=12$, 16, 24, 32 and 48. 
Those ensembles were then gauge-fixed to the Coulomb gauge by
minimizing the gauge functional 
\begin{equation}
  F_{U}[g] = \frac{1}{3} \sum_{x}
  \sum_{i=1}^{3} \Re\Tr\left(1 -g_x U_{x,i} g^{\dagger}_{x+\hat{i}} \right) \; ,
  \label{eq:func_coulomb_gauge}
\end{equation}
that involves all space-like links on the lattice. This was
accomplished by adjusting the gauge transformations $g_x \in SU(3)$ while
keeping the original gauge configuration $U$ fixed. Due to the
particular form of $F_{U}[g]$ no condition is 
imposed on time-like links. Consequently, the different time-slices can
be minimized independently. We considered gauge-fixing within a given
time-slice successful as soon as the stopping criterion
\begin{equation}
  \max_{\vec{x},\ t\,\text{fix}} \Tr \left[ \left(\partial_{i} \,^{g}\!
      A_{x,i} \right) \left(\partial_{i} \,^{g}\! A_{x,i} \right)^{\dagger}
  \right] < 10^{-13}
  \label{eq:local_gauge_violation}
\end{equation}
was satisfied. Here the lattice gauge-potential is defined in the
usual way as
\begin{equation}
  ^g\!A_{x+\hat{i}/2,i} = \frac{1}{2iag_{0}} \Big( \,^gU_{x,i} - \,
  ^gU^{\dagger}_{x,i} \Big)\Big|_{traceless} \; ,
\label{eq:transversal_potential}
\end{equation}
where $^gU_{x,i} \equiv g_x U_{x,i} g^{\dagger}_{x+\hat{i}}\;$, $a$ is
the lattice spacing and $g_0$ the bare coupling constant which
is related to $\beta$ through $\beta=6/g^2_0$.

To minimize the gauge functional we used an over-relaxation (OR) algorithm 
preceded by an optimally-tuned simulated annealing (SA) algorithm. 
In what follows, we call this particular combination of simulated annealing and
over-relaxation steps the SA-OR algorithm. To assess the influence of 
Gribov copies, we also generated some gauge copies with the pure OR
algorithm without preconditioning. In all cases, the over-relaxation 
parameter was tuned to $\omega=1.70$ on the small and
$\omega=1.60$ on the large lattices. More details are given below.

The SA algorithm has been proven to be very useful in handling various
optimization problems. For this algorithm the gauge functional $F_{U}[g]$ is
regarded as a ``spin Hamiltonian'' where the gauge transformation
fields $g_x$ take the role of ``spin variables'' coupled through the links
$U_{x,i}$ (kept fixed). Minimizing $F_{U}[g]$ is achieved by
adiabatically lowering the auxiliary temperature $T$ of a statistical
spin glass system characterized by the Gibbs weight
\begin{equation}
  W[g] \propto \exp\left(-F_U[g]/T\right) \; .
  \label{eq:spin-model}
\end{equation}
The minimization process always starts with equilibrating this spin
system at some initial temperature $T=T_i$ which is then slowly decreased. 
Formally, in the limit of (adiabatically) lowering $T \to 0$ this system
approaches the ground state and hence the gauge functional reaches its
absolute minimum for a given gauge configuration.
For the practical purpose considered here such an adiabatic cooling-down process
is not feasible as it would require an enormous amount of computing time. 
Nevertheless, we find that much lower minima for $F_{U}[g]$ can be reached, 
compared to applying only over-relaxation (OR), if we combine the SA with
the OR algorithm as follows: We start from an initial temperature of
$T_{i}=0.45$ and linearly decrease the temperature down to $T_{f}=0.01$
within 1500 ``compound sweeps''. Each such sweep consists of one
heatbath and three microcanonical update sweeps. After this, we use the OR
algorithm until the Coulomb gauge is reached, i.e.\ the
stopping criterion (\ref{eq:local_gauge_violation}) is satisfied.

The advantage of using the SA-OR instead of the OR algorithm alone
becomes more pronounced as the lattice becomes larger. Furthermore,
the number of necessary iterations in the subsequent OR algorithm is
drastically reduced by a preceding SA algorithm, the more the lower the
final $T_f$ is chosen. Note that instead
of adding subsequent OR steps, we could also have used SA on its own
extending its use to a much lower temperature $T_{f}$ to fix to Coulomb gauge. 
This, however, is much more CPU-time intensive and we find no benefit
in doing this, because after gauge-fixing the
transversality condition (\ref{eq:local_gauge_violation}) must at any case
be guaranteed with high precision, which can be achieved only by the finalizing OR.

We observe that the time-slices of a given
configuration may behave very differently during the iterative
gauge-fixing process. In fact, we find the number of necessary
iterations may differ by a factor of 10 to 20 between the individual
time-slices of a given configuration. In the majority of cases,
time-slices did not show any recalcitrancy during gauge fixing,
although in some cases time-slices could not be fixed within a
certain (predefined) number of iterations. In the latter case we simply repeated
the entire gauge-fixing process for {\it these} time-slices, using the same
algorithm but starting from a different randomly chosen gauge
transformation. The ``well-behaved'' and hence already gauge-fixed 
time-slices were not touched again. 

After all individual time-slices had been minimized, 
the original configuration $U$ was gauge-transformed, i.e.,
$U_{x,\mu}\to{}^gU_{x,\mu}$. To simplify the notation we drop the label
$g$ in what follows and assume that a gauge configuration $U$
satisfies the Coulomb gauge condition already. Since our observables,
namely the effective Coulomb potential and the ghost propagator, are genuine
three-dimensional, instantaneous observables defined by space-like links only, 
we did not have to fix the residual gauge freedom.
The latter, after the Coulomb gauge has been fixed, resides in spatially constant 
but time-dependent gauge transformations (for a continuum view at this problem
see \cite{Watson:2007fm}).

\subsection{Observables of interest}

The Coulomb energy is a complicated functional of the
transverse gauge potential $A_i({\vec x})$ and the total color charge
density. Nevertheless, it is instructive to characterize its gross
features through the infrared and ultraviolet behavior of the
expectation value of the color-diagonal part of the kernel
$M^{-1} (-\triangle) M^{-1}$ in momentum space alone. On the
lattice this is defined as the MC average
\begin{equation}
V^{L}_{\coul}({\vec k}) = \frac{1}{8 L_s^3} \left\langle
\sum_{a,{\vec x},{\vec y}} e^{i{\vec k} \cdot ({\vec x}-{\vec y})}
\Big[M^{-1}(-\triangle)M^{-1}\Big]^{aa}_{\vec{x}\vec{y}} \right\rangle \; ,
\label{eq:potential_in_momentumspace}
\end{equation}
where we use a shorthand notation for the scalar product
${\vec{k}\cdot\vec{x}} = 2\pi \sum_{i=1}^3 k_{i}x_{i}/L_i$ with
integer-valued lattice momenta $k_i$ and lattice coordinates $x_i$. 
$M$ is the lattice Faddeev-Popov operator for the Coulomb gauge 
\begin{align}
  M^{ab}_{xy} =  \delta_{x_4,y_4} &\sum_{i=1}^{3}
  \Re\Tr\Big[\left\{T^a,T^b\right\} \left(U_{x,i} +
  U_{x-\hat{i},i}\right) \delta_{{\vec x},{\vec y}} \nonumber
  \\ &\hspace{-3em}-2\,T^bT^a\, U_{x,i}\, \delta_{\vec{x}+\hat{i},\vec{y}}
  - 2\,T^aT^b\, U_{x-\hat{i},i}\,\delta_{\vec{x}-\hat{i},\vec{y}}
  \Big] \; .
  \label{eq:FP_operator}
\end{align}
Note that the Faddeev-Popov operator is a direct sum of operators
acting within individual time-slices. In coordinate space these 
three-dimensional operators define the Coulomb energy of a given
dynamical (gluonic) color-charge density plus an external one
(cf.~\Eq{eq:Coulomb_energy}). Given the tree-level form of the
Coulomb potential on a three-dimensional lattice we relate 
integer-valued lattice momenta $k_i\in (-L_i/2,L_i/2]$ to physical
ones by
\begin{equation}
  q_{i}(k_{i}) = \frac{2}{a} \sin\left(\frac{\pi k_{i}}{L_i}\right) \; .
  \label{eq:physical_momenta}
\end{equation}
Physical units are assigned by using the interpolation formula for
$r_0/a$ as given in \cite{Necco:2001xg} setting $r_0=0.5~\text{fm}$. To
simplify the writing we introduce $q$ as abbreviation for
$|\vec{q}\,|$ whenever appropriate. 

In Ref.~\cite{Zwanziger:2003de} an analytic calculation of the Coulomb
potential is presented which reads, upon Fourier transformation,
\begin{equation}
  V_{\coul}(q) = q^2 G^2(q) + V^{\text{c}}(q) \; .
  \label{eq:cp_factorisation}
\end{equation}
Here $G$ denotes the ghost propagator (entering the disconnected part)
and $V^{\text{c}}$ denotes the connected part of the potential. Under the 
assumption that the (yet unknown) connected part can be neglected, an infrared
asymptotic limit for $V_{\coul}$ has been given in \cite{Zwanziger:2003de}.
It will be analyzed below at what momenta the factorization
$V_{\coul}(q) \simeq q^2 G^2(q)$ is justified from our data concerning both
the effective Coulomb potential and the ghost propagator. The latter
can be estimated in momentum space as the MC average 
\begin{equation}
  G^{L}\big(\vec{k}\big) = \frac{1}{8 L_s^3} \left\langle
    \sum_{a,\vec{x},\vec{y}} e^{i\vec{k} \cdot (\vec{x}-\vec{y})}
    \big[M^{-1}\big]^{aa}_{\vec{x}\vec{y}} \right\rangle
\label{eq:ghostprop_in_momentumspace}
\end{equation}
at non-zero lattice momenta $k$. As for the Coulomb potential we use
\Eq{eq:physical_momenta} to assign physical momenta to $G$. To invert the
Faddeev-Popov operator we adapted the techniques
developed in Landau-gauge studies of the ghost propagator 
(see, e.g.,  \cite{Sternbeck:2005tk}).
The data for the ghost propagator used to test the factorization hypothesis
will not be presented in the present publication. They have been presented
at Lattice 2007~\cite{Voigt:2007wd} and will be discussed more in depth 
in a forthcoming paper~\cite{BerlinOsaka:2008}.

Note that both the evaluation of the effective Coulomb potential and 
of the ghost propagator involve CPU-time intensive operations. As a
consequence, we have restricted our measurements to lattice momenta
$k$ that survive a cylinder cut. Our cylinder cut is the obvious
adaptation of the Landau-gauge cylinder cut~\cite{Leinweber:1998uu}. 
To minimize finite-volumes effects, we also cone cut our data
\cite{Leinweber:1998uu} if they refer to lattices smaller than $(2.5~\text{fm})^4$.

\subsection{Running coupling and physical scale}
\label{sec:running_coupling}

The Coulomb potential is a renormalization-group 
invariant which can be written as (here for pure SU(3) gauge
theory)~\cite{Cucchieri:2000hv} 
\begin{equation}
  q^2 V_{\coul}(q) = \frac{12}{11}~g_{\coul}^2(q/\Lambda_{\coul}) \; ,
\end{equation}
where $\Lambda_{\coul}$ is a special QCD scale parameter characteristic of
the Coulomb gauge, that defines a running coupling constant $g_{\coul}$.
The latter has to satisfy the renormalization-group equation
\begin{equation}
  \label{eq:rge}
  q\frac{\partial g_{\coul}}{\partial q}=\beta_{\coul}(g_{\coul}) \; ,
\end{equation}
where the beta function, $\beta_{\coul}$, has the usual weak-coupling
expansion starting with the two standard scheme-independent coefficients
\begin{equation}
 \label{eq:b0_b1}
  b_0 \equiv \frac{11}{16\pi^2} \quad\textrm{and}\quad b_1 \equiv
  \frac{51}{128\pi^4}
\end{equation}
(see \cite{Cucchieri:2000hv} for higher terms). 
For sufficiently large
$q$, the product $11q^2V_{\coul}(q)/12$ 
is expected to be described through the two-loop
expression of the running coupling
\begin{equation}
\label{eq:alpha_s_twoloop}
  g_{\coul}^2(q) =
     \frac{1}{b_0\ln(q^2/\Lambda_{\coul}^2)}\Bigg[
     1 - \frac{b_1}{b_0^2}
     \frac{\ln[\ln(q^2/\Lambda_{\coul}^2)]}{\ln(q^2/\Lambda_{\coul}^2)}
       \Bigg] \; .
\end{equation}

Lattice data describing $q^2V_{\coul}(q)$ do not depend on the lattice
spacing~$a$ in the asymptotic scaling region. At larger~$a$, scaling 
violations should be expected though, and they will be discussed below 
for the lattice spacings used in this study.

In a previous analysis of the data~\cite{Voigt:2007wd} we used an 
ultraviolet fit to the one-loop expression 
(cf.\ the first term of \Eq{eq:alpha_s_twoloop}) to fix the unknown
physical scale of the effective Coulomb potential (see
Ref.~\cite{Voigt:2007wd} for details). For the present study, we
scrutinized if the highest momenta accessible in our simulations
really permit a feasible fit to the one-loop or the two-loop expression
given in \Eq{eq:alpha_s_twoloop}. We find that this is not the case and 
that the ultraviolet fit described in~\cite{Voigt:2007wd} has artificially 
up-scaled our data by a free factor bigger than one.
In the present study, we therefore do not rely anymore on this ultraviolet fit. 

Indeed, the physical scale is fixed by
simply multiplying the bare lattice data for the effective Coulomb
potential with $6/(\beta a^{2})$
\begin{equation}
V_{\coul}(q) = \frac{6}{\beta} a^{2} V^{L}_{\coul}(k,\beta) \; ,
\end{equation}
where $a$ denotes the lattice spacing in $\mbox{GeV}^{-1}$. 
Again, we use the interpolation formula in \cite{Necco:2001xg}
to set $a$ assuming $r_0=0.5\ \text{fm}$. For all figures in the present
paper, the physical scale of the effective Coulomb potential is fixed in 
this way.

\section{Studying systematic effects}
\label{sec:systematics}

\begin{figure*}
  \includegraphics[width=1.0\textwidth]{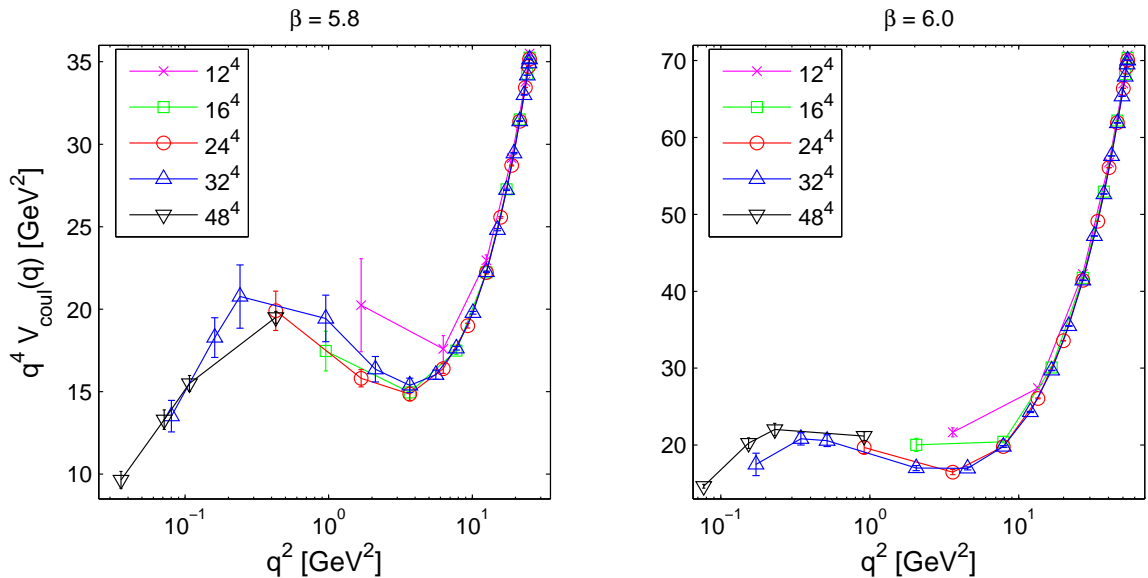}
  \caption{(Color online) The Coulomb potential multiplied by $q^4$ 
    shown as a function of $q^2$ in physical units. We show data for 
    different lattice sizes at $\beta=5.8$ (left) and $\beta=6.0$ (right) 
    to illustrate finite-volume effects. These seem to be under control
    for data on lattices larger than $16^4$ because those fall roughly 
    on the same curve. For both $\beta$ values we only used data from 
    first SA-OR copies (fc).}
\label{fig:cp_finitevolume}
\end{figure*}

\begin{figure*}
  \includegraphics[width=1.0\textwidth]{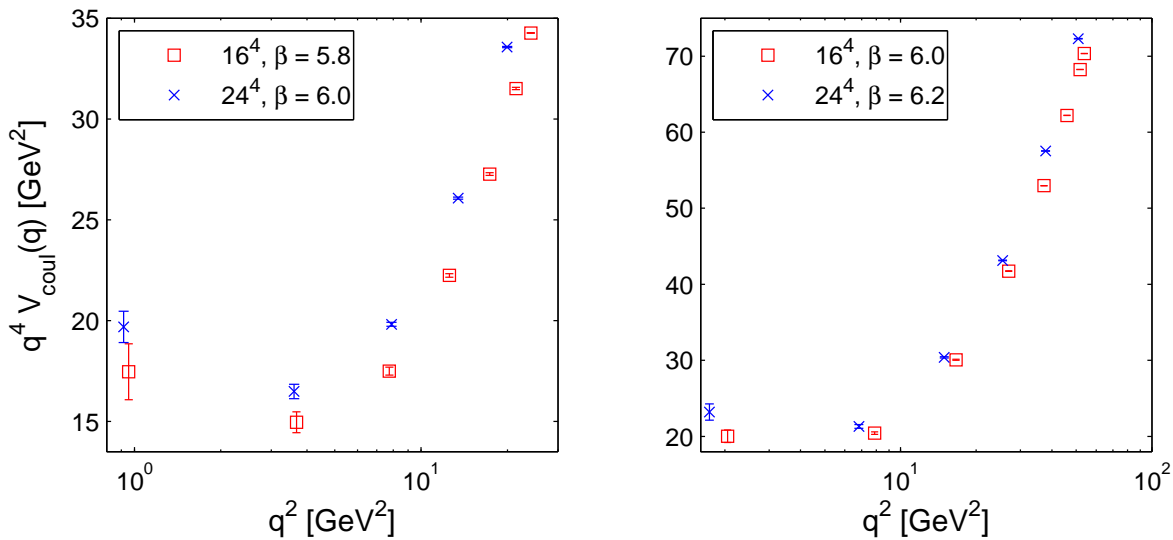}
  \caption{(Color online) The Coulomb potential multiplied by $q^4$ 
    shown versus $q^2$ measured on comparable physical volumes. 
    Data for $\beta=5.8$ and $6.0$ are shown on the left,
    and for $\beta=6.0$ and $6.2$ on the right hand side. 
    Only data from first SA-OR copies is shown.} 
\label{fig:cp_errorinbeta}
\end{figure*}

In this section we discuss the effects of finite lattice volumes 
and lattice spacings as well as the influence of the Gribov ambiguity on the
effective Coulomb potential. 

\subsection{Lattice artifacts}

As we are primarily interested
in the product $q^4~V_{\coul}(q)$, we directly discuss this product
instead of the effective Coulomb potential itself. 
Note that we investigate effects of finite lattice spacings and
volumes by considering Coulomb-potential data collected for first
SA-OR copies only.

Finite-volume effects are studied by varying the lattice sizes
from $12^4$ to $48^4$ but keeping $\beta$, and hence the lattice
spacing $a$, fixed. We find that only data obtained on the smaller
lattices, $12^4$ and $16^4$, at $\beta=6.0$ show visible finite-volume
effects at lower momenta. For larger lattices, namely $24^4$,
$32^4$ and $48^4$, effects seem to be mild
(see the magnified view in the right panel of \Fig{fig:cp_finitevolume}). 
At $\beta=5.8$ (reaching even lower momenta) only the $12^4$ data
clearly deviate from the other data (see \Fig{fig:cp_finitevolume},
left panel).

Lattice-spacing effects are investigated by comparing
data from lattices of equal physical volume for different values of
$\beta$. Within our choice of $\beta$ values and lattice sizes $L^4$,
we can find only a few combinations of $\beta$ and $L$ where
this is approximately possible. For those we can compare data at
approximately equal physical momenta and disentangle by eye the effect
of varying $a$. As demonstrated in \Fig{fig:cp_errorinbeta} these
discretization effects are small and of the order of 10 to 15\% at
largest. The difference between data for $\beta=6.2$ and 6.0 is
smaller (right panel) than the difference between data for $\beta=6.0$
and 5.8 (left panel).

\begin{figure*}
  \includegraphics[width=0.9\textwidth]{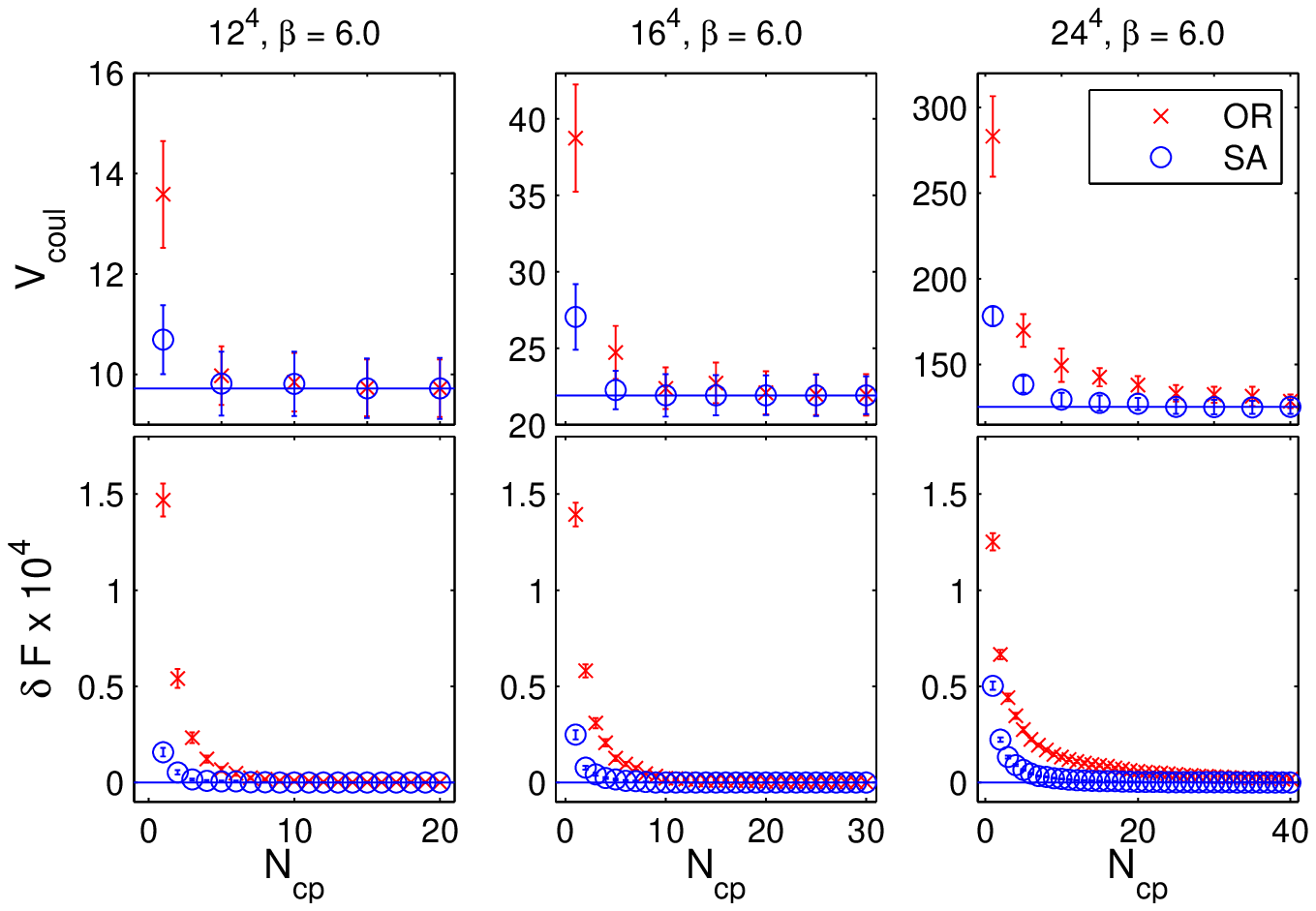}
    \caption{(Color online) The upper row shows the convergence of $V_{\coul}$ 
    with increasing number $\Ncp$ of copies obtained with either the 
    SA-OR (circles) or the OR algorithm (crosses). The Coulomb potential is 
    shown for the smallest (on-axis) lattice momentum available for each of
    the three lattice sizes, i.e. $\vec{k}=([1,0,0])$ with square brackets 
    indicating that the average over all permutations is taken.
    The lower row shows the corresponding convergence of the (average) deviation 
    of the gauge functional from its best value known for each configuration. 
    As a function of $\Ncp < \Ncp^{\mxx}$, the values of $V_{\coul}$ and 
    $F$ are to be understood as those for the ``currently best'' gauge copy 
    among the $\Ncp$ copies inspected for each configuration after $\Ncp$
    repetitions of gauge fixing.}
\label{fig:cpandfunc_convergence}
\end{figure*}

\begin{figure*}
  \centering
  \includegraphics[width=0.6\linewidth]{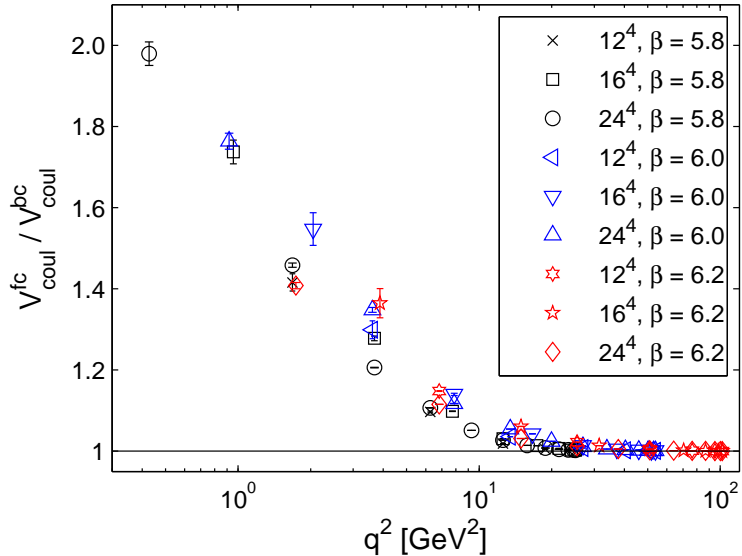}
  \caption{(Color online) The ratio of the Coulomb potential (as function of $q^2$) 
  if evaluated either on arbitrary (first) copies from simple OR or on best gauge
  copies from repeated SA-OR. Note that the enhancement for the first copy 
  tremendously grows with $q^2 \to 0$ compared to what is known in the case of the
  ghost propagator. In contrast to the upper panel of \Fig{fig:cpandfunc_convergence} 
  only momenta are included that are allowed by the cuts mentioned above.}
  \label{fig:cp_bcvsfc}
\end{figure*}

\subsection{Effects due to the Gribov ambiguity}

In comparison to lattice artifacts, the Gribov ambiguity turns out to
have a much larger impact on the Coulomb potential data. In order to
assess the influence of this ambiguity we follow the ``first copy --
best copy'' (fc-bc) strategy applied before in Landau-gauge studies of
gluon and ghost propagators \cite{Bakeev:2003rr,Sternbeck:2005tk}.
Here, we use this strategy in two different ways. 

First, we estimate the number $\Ncp$ of gauge-fixed copies per configuration
necessary to achieve ``quasi convergence'' of the Coulomb potential, considered 
as a function of $\Ncp$. Second, we quantify the systematic error of the 
Coulomb potential that is admitted if {\it an arbitrary} (first) gauge-fixed copy 
is chosen instead of {\it the best} copy among $\Ncp$ copies for each gauge 
configuration. A copy is considered to be the best among all $\Ncp$ gauge-fixed 
copies of a given configuration if its gauge functional is lower than 
those of all the other $\Ncp-1$ copies after the gauge-fixing has been attempted
$\Ncp$-times. 

Let us first compare the convergence of the bare data describing the Coulomb
potential upon increasing $\Ncp \to \Ncp^{\mxx}$ for the two gauge-fixing
algorithms OR and SA-OR. As an example, in \Fig{fig:cpandfunc_convergence} we 
show for each lattice size data describing $V_{\coul}$ for the lowest (on-axis)
lattice momentum available, i.e. $\vec{k}=([1,0,0])$ with square brackets 
indicating that the average over all permutations is taken.
Note that in contrast to the rest of this paper we did not apply neither the 
cylinder nor the cone cut here. The obtained deviations from the best-copy value 
can be considered as an upper bound for all other momenta.

The data were obtained at $\beta=6.0$ on lattice sizes $L^4=12^4,16^4$ and $24^4$.
For each gauge-field configuration a number of $\Ncp^{\mxx}=20$, 30 and 40 independent 
gauge-fixed copies was generated separately with the OR and the SA-OR algorithm. 
From the figure we see that upon increasing $\Ncp$ the effective Coulomb potential 
decreases and becomes (more or less) independent of $\Ncp$ for $\Ncp$ coming 
closer to $\Ncp^{\mxx}$. What $\Ncp$ is sufficient to achieve ``quasi convergence'' 
depends, of course, on the gauge-fixing algorithm and on the lattice size. In fact, 
it is clearly visible in \Fig{fig:cpandfunc_convergence} that the number of
gauge copies necessary to achieve convergence is substantially lower
for the SA-OR algorithm than for the OR algorithm. With both algorithms one needs 
to consider more gauge copies with growing $L$. 
For example, if we use the SA-OR algorithm, a number of copies $\Ncp=5$  
on a $12^4$ lattice and $\Ncp=15$ on a $24^4$ lattice is sufficient. In contrast, 
if we were using the OR algorithm, $\Ncp=40$ or more copies are necessary for a 
lattice like $24^4$. Note that the observed increase with $L$ at fixed $\beta$ is 
partly due to the smaller physical value associated with the lowest on-axis lattice
momentum that needs to be considered with increasing $L$. 

On the larger lattices $32^4$ and $48^4$ we could not afford to gauge-fix more 
than a single gauge copy per configuration with the SA-OR algorithm. This was 
simply due to a drastic increase of the necessary number of iterations, but also 
due to a growing number of ``trouble-making'' time-slices encountered on those 
larger lattices. Thus, we did not apply the OR algorithm for the purpose of 
comparison, and therefore we are not in the position to assess the influence of Gribov 
copies on these lattices at the present stage. 

As mentioned above, we also used the fc-bc strategy to estimate the impact of 
Gribov-copy effects on the Coulomb potential data at different physical momenta. 
For this purpose we gauge-fixed our field configurations at $\beta=5.8$, $6.0$ and $6.2$ 
to Coulomb gauge only once with the OR algorithm on one hand and $\Ncp$-times
with the SA-OR algorithm on the other. In order to obtain the results shown in
\Fig{fig:cp_bcvsfc} we have chosen, refering to \Fig{fig:cpandfunc_convergence},
$\Ncp=10$, 15 and 20 as sufficient numbers of gauge-fixed copies per configuration 
for the lattice sizes $12^4$, $16^4$ and $24^4$, respectively. Then the Coulomb 
potential was measured separately on the set of single copies obtained with the
OR algorithm on one hand and on the set of best copies obtained with the repeated
SA-OR algorithm on the other. For brevity we refer below to these two sets as the 
first OR and the best SA-OR copies. 

The ratio of the effective Coulomb potential measured for first OR copies and
best SA-OR copies is depicted in \Fig{fig:cp_bcvsfc} as a function of momentum
squared. The Gribov ambiguity has a dramatic impact on the effective Coulomb
potential at $q^2 < 10~\text{GeV}^2$. Even for the rather
small lattices considered here, and hence for rather high physical
momenta, the measurement of the effective Coulomb potential on first OR-copies 
gives results larger by up to 100\% than the results on best SA-OR copies. 
Note that this effect is much stronger than what has been observed for the ghost
propagator using the same method \cite{Voigt:2007wd,BerlinOsaka:2008}. 
There an enhancement of about 5 to 10\% was typical for the presently 
accessible lowest momenta. Note also that here we are comparing the 
standard with one of the best presently known methods of gauge fixing.

In order to assess next the difference between $V_{\coul}$ obtained from an arbitrary,
first SA-OR copy (fc) and the best among a sufficient number of SA-OR copies (bc),
restricted, however, to the smallest lattice momentum for each lattice size, i.e.
${\vec k} = ([1,0,0])$, we have to look back to \Fig{fig:cpandfunc_convergence}.
Considering the ratio $R$ of the effective Coulomb potential measured either for 
first or best SA-OR copies as a function of the lattice size $L$, we find that this 
is well described by $R \approx c - d/L$. If such an ansatz was used to extrapolate 
the ratio $R$ at $\beta=6.0$ to $L=48$, a ratio $R=1.6$ would be obtained. For
the first OR copy an overestimation factor $R=2.7$ would be expected.
Both are reasonable upper bounds for the overestimation of the Coulomb potential
at any fixed physical momentum for first SA-OR copies and - even worse - first 
OR copies. This estimate will be needed in the next Sect.~\ref{sec:infrared}.

We conclude that the effective Coulomb potential is less affected by Gribov-copy 
effects if we use the SA-OR algorithm instead of the OR algorithm.
This conclusion rests on the observation that the ``quasi convergence'' 
of the Coulomb potential is faster for the SA-OR than for the OR algorithm. 
Second, the results obtained on arbitrary, first SA-OR copies (fc) are less affected 
by the Gribov ambiguity than those obtained on arbitrary, first OR-copies. 
Therefore, we have used the SA-OR algorithm as our method of choice for the results 
to be presented in the following. Recall, however, that if only first SA-OR copies 
are available for analysis, measurements of the Coulomb potential in the infrared
region will be accompanied with an increased uncertainty. For instance, an overestimation 
of about $60 \%$ for the smallest lattice momentum on a $48^4$ lattice at $\beta=6.0$
must be expected.
     
\section{Infrared behavior}
\label{sec:infrared}

Despite the Gribov ambiguity being that large, we now try to summarize what we know 
about the momentum dependence of the effective Coulomb potential, globally and in 
particular in the low-momentum region. As mentioned above, we were not in the position 
to generate more than a single SA-OR gauge copy per configuration on the larger 
lattices $32^4$ and $48^4$.
Therefore we present here a full set of data concerning the Coulomb potential for
a single SA-OR copy (fc) per configuration for all $\beta$ values and lattices sizes,
ensuring in this way an equal treatment of Gribov copy effects on both small and large
lattices. As is well known, this choice is equivalent to an averaging over all local 
minima of all configurations, i.e. all over the {\it Gribov Region}. Best-copy data, 
that we have available only up to lattices $24^4$ (after inspecting a sufficient number 
$\Ncp$ of copies) would come close to a prescription that requires an average over
only the absolute minimum per configuration, i.e. restricted to the {\it Fundamental
Modular Region}. Zwanziger has argued that these averages should approach each other 
in the limit of large volumes.    

\begin{figure*}
  \includegraphics[width=1.0\textwidth]{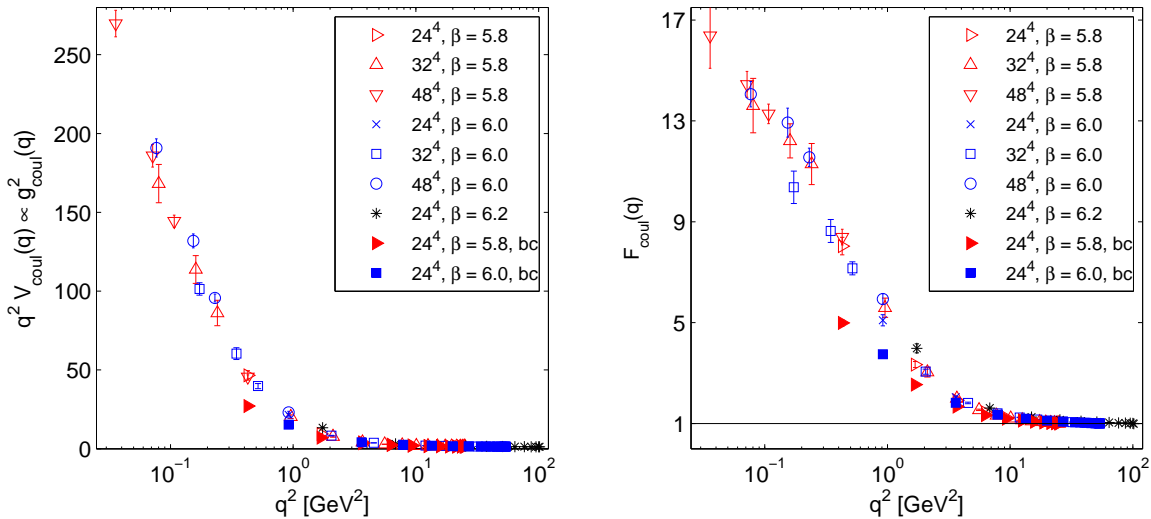}
  \caption{(Color online) Left: The running coupling $g^2_{\coul}(q) \propto q^2 V_{\coul}(q)$ 
   diverges in the infrared region and tends to zero in the asymptotic limit 
   $q^2 \rightarrow \infty$. 
   Right: The factorization of the effective Coulomb potential is violated for momenta 
   $0.04 \mbox{ GeV}^2 \leq q^2 \leq 10 \mbox{ GeV}^2$. For both the running coupling 
   and the test of the factorization hypothesis, the open symbols (including stars) 
   represent measurements on the first SA-OR copies per configuration. For comparison, 
   the filled triangles and filled squares show selected results for the best SA-OR 
   copy (bc) per configuration for two $\beta$ values on the largest lattice $24^4$ 
   where the Gribov problem was fully under control.}
  \label{fig:cp_runningcoupling}
\end{figure*}

\begin{figure*}
\includegraphics[width=1.0\linewidth]{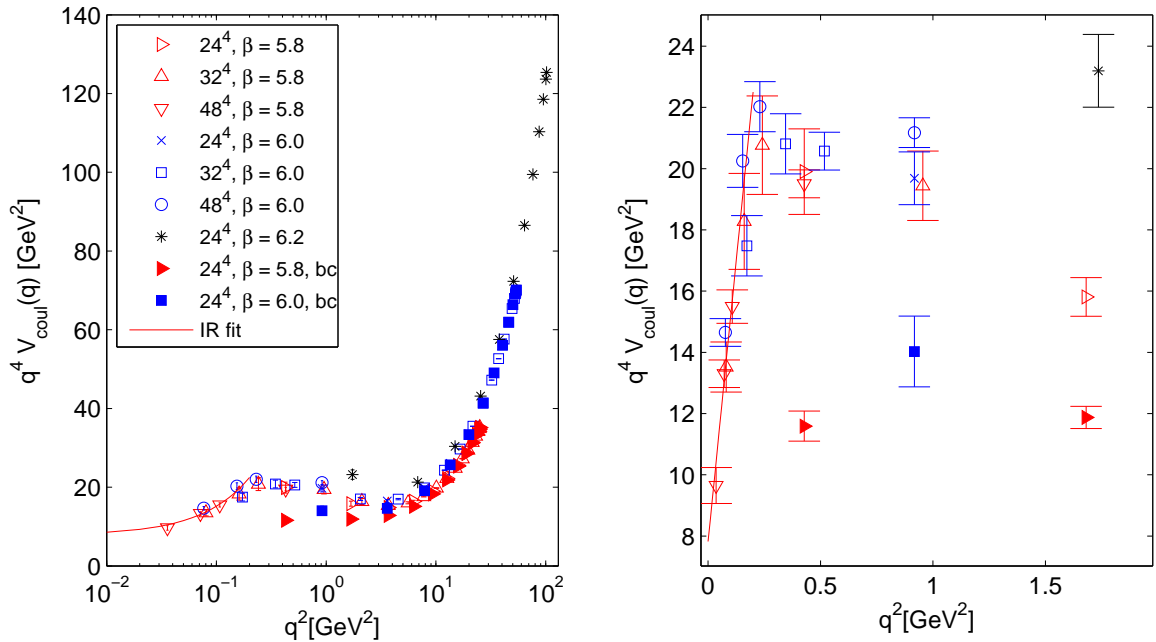}
\caption{(Color online) The effective Coulomb potential multiplied by $q^4$ as a 
function of the physical momentum squared. Left: with a logarithmic momentum scale
in order to overlook both the IR and UV behavior. Right: the infrared momentum 
region is shown in a linear scale in $q^2$ in order to judge the adequacy of the 
linear fit of the first-copy data in the extremely IR region. The infrared fit 
used to extract the corresponding Coulomb string tension is also shown in the 
left panel. Open symbols (including stars) represent measurements on the first 
SA-OR copies per configuration. For comparison, the filled triangles and filled 
squares in both panels show results for the best SA-OR gauge copies (bc) for two 
$\beta$ values on the lattice $24^4$, the largest lattice where the Gribov problem 
was under scrutiny.}
\label{fig:cp_stringtension}
\end{figure*}

As long as they did not converge, we are admitting a strong systematic effect when
we restrict the analysis to the first SA-OR copy.
This can be clearly seen in our data from smaller lattices, $12^4$, $16^4$ and $24^4$.
In order not to overload the \Fig{fig:cp_runningcoupling} we show here 
(and lateron in \Fig{fig:cp_stringtension}) only {\it selected}
results from SA-OR best-copies. The data are from the $24^4$ lattice where we had
the choice between 40 copies at $\beta=6.0$ and between 20 copies at $\beta=5.8$.
We try here (and later for the Coulomb potential) our best to estimate the systematic
error emerging from the ignorance of further Gribov copies at larger lattices.
A thorough study of the Gribov ambiguity for larger lattices remains highly desirable.

In the infrared momentum region, the running coupling given through the effective
Coulomb potential diverges stronger than $1/q^2$. This is shown in the left panel of 
\Fig{fig:cp_runningcoupling}. The very fact of the infrared enhancement will not need to 
be revised if once a systematic account for the Gribov effect would be undertaken, 
although the divergence would be less pronounced. A rough indication of the size of 
the effect is given by the filled symbols in that figure. These are best-copy results 
for the lattice $24^4$.

In Ref.~\cite{Zwanziger:2003de}, Zwanziger presented an analytic calculation of the 
Coulomb potential. By only considering the disconnected part of the expectation value 
of the effective Coulomb potential (cf.~\Eq{eq:cp_factorisation}) Zwanziger predicted 
an almost linearly rising effective Coulomb potential in the infrared limit. 
Using our data for the ghost propagator~\cite{Voigt:2007wd} we are now in the position
to test the validity of his factorization hypothesis. If Zwanziger's assumption were valid, 
the ratio
\begin{equation}
F_{\coul}(q) = \frac{V^L_{\coul}(q)}{(a~q~G^L(q))^{2}} \; .
\end{equation}
should be constant as function of the momenta. Note that $V^{L}$ and $G^L$ denote the bare
lattice Coulomb potential and the lattice ghost propagator taken at the
{\it physical} momentum $q$. The resulting plot shown in the right panel of 
\Fig{fig:cp_runningcoupling} demonstrates that the assumption of factorization is valid 
only for $q^2 > 10 \mbox{~GeV}^2$, but it is not correct in the momentum range 
$q^2 \leq 10 \mbox{~GeV}^2$. This is in agreement with the results 
of Langfeld and Moyaerts for $SU(2)$ pure gauge theory~\cite{Langfeld:2004qs}.

Much alike the enhancement of the running coupling, our conclusion that 
the factorization hypothesis is violated would also not be invalidated if the effect of Gribov
copies was taken into account properly.
Similar to the left panel, the anticipated Gribov effect on the violation of factorization
is shown by the filled symbols in the right panel, which are the best-copy results for the
lattice $24^4$.    

We discuss now the momentum dependence of the effective Coulomb potential 
in three clearly emerging momentum ranges, the {\it high-momentum range}, the 
{\it intermediate momentum range} and the {\it low-momentum range}, and 
describe the influence of the Gribov ambiguity in each range separatly.

The left panel of \Fig{fig:cp_stringtension} shows the presently known picture 
concerning the momentum dependence of $q^4~V_{\coul}(q)$. A logarithmic momentum 
scale has been chosen in order to give a global view including the ultraviolet 
and infrared behavior. 
For the largest momenta in the {\it high-momentum range} $q^2 \geq 10 \mbox{~GeV}^2$,
the Coulomb potential shows roughly the expected $1/q^2$ behavior leading to an increase 
of $q^4~V_{\coul}(q)$ linear in $q^2$. From \Fig{fig:cp_bcvsfc} it is clear that the 
high-momentum region is robust with respect to the Gribov ambiguity.
Although the inspection by eye suggests that we are seeing the tree-level form of the 
Coulomb potential, we could not find reasonable fits of our data by the one-loop or the two-loop
expressions given in \Eq{eq:alpha_s_twoloop}.
We conclude that much higher momenta must be considered to get an estimate of the Coulomb
scale parameter $\Lambda_{\coul}$ from such a fit.

With decreasing physical momenta, the first-copy data for $q^4~V_{\coul}(q)$ reach 
an almost flat region in the {\it intermediate momentum range}
$ 0.2 \mbox{~GeV}^2 \leq q^2 \leq 6 \mbox{~GeV}^2$, although a little bulge is visible
in the left panel of \Fig{fig:cp_stringtension}.   
If the function $q^4~V_{\coul}(q)$ stayed constant on the level of $\approx 20 \mbox{~GeV}^2$
in the limit $q^2 \to 0$, this would imply a perfect linearly confining potential
corresponding to an estimate of $\sigma_{\coul} \approx (890 \mbox{~MeV})^2$. 
This figure is more likely an upper bound. 

Indeed, if for large spatial distances $r$ we assume the simple ansatz~\cite{Cucchieri:2002su}
\begin{equation}
V_{\coul}(r) = - \sigma_{\coul}\,r + C/r \; ,
\label{eq:cp_infraredpot}
\end{equation}
this suggests a momentum behavior
\begin{equation}
 V_{\coul}(q) = \frac{8\pi \sigma_{\coul}}{q^{4}} + \frac{4\pi C}{q^{2}} \; ,
\label{eq:cp_infraredfit}
\end{equation}
with the intercept of $q^4~V_{\coul}(q)$ at $q^2=0$ defining the Coulomb string tension $\sigma_{\coul}$.

In the {\it intermediate momentum range} $ 0.2 \mbox{~GeV}^2 \leq q^2 \leq 6 \mbox{~GeV}^2$ 
the Gribov effect sets in and becomes apparently more severe with decreasing momentum.
For instance, for the smallest (on-axis) lattice momenta on lattices of sizes $12^4$,
$16^4$ and $24^4$ we have seen in \Fig{fig:cpandfunc_convergence} that the Coulomb potential
$V_{\coul}(q)$ is overestimated by the first SA-OR copies compared with the best SA-OR copies
(among 40 copies). For the $24^4$ lattice at $\beta=6.0$ the overestimation amounts to 
$\approx 40\%$. This can be extrapolated to the $48^4$ lattice where the effect amounts to
$\approx 60 \%$. This is an upper bound for the Gribov effect experienced by $V_{\coul}(q)$ 
at {\it physical} momenta that are allowed by the cylinder and cone cuts.

In agreement with these estimates it can be seen in the left panel of
\Fig{fig:cp_stringtension} that in the {\it intermediate momentum range} 
$ 0.2 \mbox{~GeV}^2 \leq q^2 \leq 6 \mbox{~GeV}^2$ the best-copy data from
the $24^4$ lattice (shown as filled symbols) provide us with another, independent 
early indication of a plateau. The somewhat lower level of $\approx 10 \mbox{~GeV}^2$ 
would correspond to $\sigma_{\coul} \approx (630 \mbox{~MeV})^2$. In view of this the 
bulge must be understood as an artefact of insufficient gauge fixing.

With the simulation reported here, on our largest lattices the {\it low-momentum range} 
with $q^2 < 0.2$ has become accessible for the first time.
Rather unexpectedly in this region the first-copy data for $q^4~V_{\coul}(q)$ drop
with decreasing momentum as seen in the left panel. The right panel of 
\Fig{fig:cp_stringtension} shows the infrared region magnified and in a linear scale 
in $q^2$. This picture shows that a fit ansatz linear in $q^2$ describes the drop of 
the first-copy data very well. We do not know whether a similar effect, namely the 
onset of an apparently new infrared regime in the {\it low-momentum range} will happen 
for the best-copy data as well. For the time being we assume that the bulge and the 
new infrared regime is only a matter of measurements on insufficiently gauge-fixed 
configurations. We have fitted the behavior according to \Eq{eq:cp_infraredfit}. 
The Coulomb string tension is estimated as 
\begin{equation}
\sigma_{\coul} = (552 \pm 35 \mbox{~MeV})^2 \; .
\end{equation}
With some caution we may consider this as the {\it common limit} for $q^2 \to 0$ 
and the {\it common lower bound} for the Coulomb string tension (common to both
standards of gauge-fixing).

The other fit parameter, the ``Coulombic'' coefficient $C$ in front of the $1/r$ term
in \Eq{eq:cp_infraredpot}, is obtained as 
\begin{equation}
C = 6.0 \pm 1.0  \; .
\end{equation}
This parameter has no relation to the ``Coulombic'' part $1/r$ 
in \Eq{eq:cp_infraredpot}. It rather describes the narrow momentum interval where the
single-copy data probably converge to the best-copy results for $q^4~V_{\coul}(q)$. 
The small number of data points is another reason why we give not much significance 
to the fit. Still, details of the least $\chi^2$ fit of the first-copy data are
shown in \Tab{tab:infraredfit}.
\begin{table}
\caption{Results of $\chi^2$ fits to the single-copy data at momenta $q^2\le q_{\mxx}^2$.}
\label{tab:infraredfit}
\begin{tabular}{ l@{\quad} c@{\quad} c@{\quad} c@{\quad} r }
\hline\hline
$q_{\mxx}^2 \mbox{ [GeV}^2\mbox{]}$ & \# data & $\sqrt{\sigma_{\coul}} \mbox{ [MeV]} $  
& $C$ & $\chi^2/\mbox{ndf}$ \\
& points & & & \\ \hline
   0.11 & 5 & $534(16)$  & 6.6(3) & 2.9  \\
   0.16 & 6 & $526(18)$  & 6.8(4) & 1.8  \\
   0.17 & 7 & $558(20)$  & 5.8(2) & 2.5  \\
   0.18 & 8 & $587(28)$  & 4.9(2) & 3.8  \\ \hline\hline
\end{tabular} 
\end{table}
We remark that the choice of the upper momentum cutoff for the fitting range, $q_{\mxx}$, 
has only weak influence on the fit results.

In units of the Wilson string tension the fit result is 
\begin{equation}
\sigma_{\coul} = (1.6 \pm 0.2 ) ~\sigma_{Wilson} \; .
\end{equation}
This is the tentative lower bound for the Coulomb string tension.

Our estimate for the Coulomb string tension is in agreement with Zwanziger's inequality. 
The relevance of this estimate is, however, faced with three sources of uncertainty. 
\begin{itemize}
\item First, it relies only on first-copy data in a rather small number of data points, 
and the obtained $\chi^2/\mbox{ndf}$ values are rather large (see \Tab{tab:infraredfit}).
The latter might be interpreted as a probable inadequacy of the assumed infrared ansatz
\Eq{eq:cp_infraredfit}. On the other hand, the right panel of \Fig{fig:cp_stringtension} 
supports such a behavior.
\item Second, the weak but visible scaling violation of the effective Coulomb potential has the 
effect that our estimate of the Coulomb string tension would be higher if we considered 
higher inverse coupling constants $\beta$. The effective Coulomb potential in general 
slightly increases with increasing $\beta$. 
\item Third, the strong Gribov effect is neglected 
in this estimate for the Coulomb string tension. If we had consequently looked for the best
SA-OR copies and had measured $q^4~V_{\coul}(q)$ for these, the amount of overestimation
by the first-copy data in the bulge region is of the estimated order. One possibility
is that by the drop described by the fit given above the (yet unknown) level of the best-copy
results is reached. However, we cannot exclude the possibility that the best-copy data 
in the {\it low-momentum range} will also enter a new infrared regime with a similar 
decrease, such that the overestimation by the first-copy data remains. In this case 
the final estimate of the Coulomb string tension would be close to the Wilson string 
tension. 
\end{itemize}

In the light of these uncertainties, we find it difficult to draw a conclusion on the
exact value of the Coulomb string tension.
Our value is larger than the values reported in previous $SU(2)$ investigations starting
also from the definition \Eq{eq:coulomb_potential}. These authors arrived at an estimate 
close to the Wilson string tension~\cite{Cucchieri:2002su,Langfeld:2004qs}.
However, the Gribov copy problem for the effective Coulomb potential was ignored in these 
studies. Furthermore, in Ref.~\cite{Cucchieri:2002su} the estimate of the Coulomb string
tension actually relies on data in the perturbative region, while the first plateau of 
$q^4~V_{\coul}(q)$ has been considered as a finite-volume effect. Such a plateau could 
be observed in Ref.~\cite{Langfeld:2004qs} but the further decrease of $q^4~V_{\coul}$ 
for even lower momenta was beyond the possibilities of this investigation. In contrast, 
$SU(3)$ studies using incomplete (partial-length) Polyakov lines, made in order to 
interpolate between the Coulomb string tension and the Wilson string tension, gave 
$\sigma_{\coul} = (2-3) \sigma_{Wilson}$ \cite{Nakamura:2005ux,Nakagawa:2006fk}. 
These studies also have neglected the problem of Gribov copies that might have affected 
the measured correlators. 

\section{Conclusions}
\label{sec:conclusion}

In this study we have attempted a thorough measurement of the effective 
Coulomb potential in $SU(3)$ lattice gauge theory. We used a broad range of
lattice sizes, $12^4 - 48^4$, to perform Monte-Carlo simulations at
the three values $\beta=5.8$, $6.0$ and $6.2$. This has allowed us to show that
finite-volume effects are hardly visible on the larger lattices and
discretization effects are modest. Additionally, the use of 
the fc-bc strategy has revealed a dramatic dependence of the Coulomb
potential on the choice of Gribov copy. 

Unfortunately, by computer resources we were forced to restrict this ``Gribov analysis''
to the smaller lattices $12^4$, $16^4$ and $24^4$. 
Thus, performing a full Gribov study up to the largest lattices still remains a highly 
desirable goal. We note that the necessity of choosing best copies versus first (and 
hence arbitrary) gauge-fixed copies is a matter of current debate (see also
\cite{Bogolubsky:2005wf,Bogolubsky:2007bw}). As another example, in a
BRST formulation, an average over all Gribov copies is taken which, on
the lattice, usually leads to the well-known Neuberger $0/0$ problem
\cite{Neuberger:1986vv, Neuberger:1986xz}. For a recent lattice BRST
formulation without this complication see \cite{vonSmekal:2007ns}. 

What can be said here with confidence is that for the effective Coulomb
potential we find an extraordinarily strong Gribov-copy effect which has
never been observed before for other observables (say, the gluon and ghost
propagators).

We see a strong violation of the factorization hypothesis for the effective 
Coulomb potential in momentum space below $q^2 \sim 5 \mbox{~GeV}^2$. For
smaller momenta the ``connected part'' of the corresponding expectation value
in \Eq{eq:cp_factorisation} is not negligible anymore. 
This spoils any simple relation between the momentum dependence of the effective Coulomb 
potential and the behavior of the ghost propagator. 

Using only one SA-OR gauge copy per configuration and hence
allowing a strong systematic Gribov effect to be included in the bargain,
we found a new infrared regime of the effective Coulomb potential. 
The first plateau of $q^4~V_{\coul}(q)$, encountered with decreasing
momenta, turns out {\it not} to represent the asymptotic behavior, 
because there is a further step-like decline for even smaller momenta. 
The size of the step is of the same order as the extrapolated difference
between single-copy and many-copy SA-OR results. Therefore, we adopted the
point of view that the ``breakthrough'' to some new infrared regime at large 
enough volume is a feature only of the single-copy data, and that some kind of
convergence (between averaging over the Gribov Region and the Fundamental Modular 
Region) is behind this observation.

Future studies shall scrutinize whether the presumed common infrared limit 
of these two schemes really exists or, alternatively, the Gribov ambiguity
persists at lower momentum for larger volumes. We estimated the Coulomb
string tension by fitting the data at the lowest momenta and found it
approximately 1.6 times larger than the Wilson string tension.
If the Gribov ambiguity persists, it is not excluded that in the -- further
delayed -- infrared limit finally $\sigma_{\coul}=\sigma_{\wilson}$ will be found.

\section{Acknowledgements}

This work is supported by the DFG under contract FOR 465
(Research Group {\it Lattice Hadron Phenomenology}), and by the
Australian Research Council.  A major part of the simulations were
done on the IBM pSeries 690 at HLRN, Germany.  We thank H.~St\"uben
for contributing parts of the code. We are grateful to Y.~Nagakwa,
A.~Nakamura, T.~Saito, L.~von Smekal and H.~Toki for inspiring discussions.

\bibliographystyle{apsrev} 
\bibliography{citations}

\end{document}